\documentclass[a4paper]{article}

\usepackage{INTERSPEECH2019}
\usepackage{subfigure} 
\usepackage{multirow}
\usepackage{mathrsfs}
\usepackage{makecell}
\usepackage{amsfonts}

\title{Self-Attention Transducers for End-to-End Speech Recognition}
\name{Zhengkun Tian$^{1,2}$, Jiangyan Yi$^{1}$, Jianhua Tao$^{1,2,3}$, Ye Bai$^{1,2}$, Zhengqi Wen$^{1}$}
\address{
	$^1$National Laboratory of Pattern Recognition, Institute of Automation, \\
	Chinese Academy of Sciences, Beijing, China\\
	$^2$School of Artificial Intelligence, University of Chinese Academy of Sciences, Beijing, China\\
	$^3$ CAS Center for Excellence in Brain Science and Intelligence Technology, Beijing, China}
\email{\{zhengkun.tian, jiangyan.yi, jhtao, ye.bai, zqwen\}@nlpr.ia.ac.cn}

\begin{document}

\maketitle
\begin{abstract}
Recurrent neural network transducers (RNN-T) have been successfully applied in end-to-end speech recognition. However, the recurrent structure makes it difficult for parallelization . In this paper, we propose a self-attention transducer (SA-T) for speech recognition. RNNs are replaced with self-attention blocks, which are powerful to model long-term dependencies inside sequences and able to be efficiently parallelized. Furthermore, a path-aware regularization is proposed to assist SA-T to learn alignments and improve the performance. Additionally, a chunk-flow mechanism is utilized to achieve online decoding. All experiments are conducted on a Mandarin Chinese dataset AISHELL-1. The results demonstrate that our proposed approach achieves a 21.3\% relative reduction in character error rate compared with the baseline RNN-T. In addition, the SA-T with chunk-flow mechanism can perform online decoding with only a little degradation of the performance.
\end{abstract}

\noindent\textbf{Index Terms}: end-to-end, self-attention transducer, path-aware regularization, chunk-flow mechanism

\section{Introduction}
Compared with traditional hybrid models, end-to-end models for speech recognition show great potential because of their elegant model structure, alignment-free attribute and simple training process. At present, all workable end-to-end models can be grouped into four categories: connectionist temporal classification (CTC) \cite{graves2006connectionist, graves2014towards,amodei2016deep}, attention-based sequence to sequence models \cite{bahdanau2014neural,vaswani2017attention, dong2018speech,Kim2017Joint}, end-to-end models using lattice-free MMI \cite{povey2016purely,hadian2018end} and recurrent neural network transducers \cite{graves2012sequence,rao2017exploring, graves2013speech,he2018streaming}. 

CTC was first proposed to convert input feature sequences to the corresponding text sequences, However, CTC cannot model the dependencies between the outputs and be optimized jointly with language model \cite{graves2012sequence}. As an extension of CTC, RNN-T introduces a language component named prediction network to model the dependencies between the outputs. And it can also optimize acoustic and language components jointly. 

The recurrent structure makes RNN-T able to model long-term dependencies. However, it also precludes parallelization, which becomes critical at longer utterance lengths. Recently, the self-attention network, has been successfully applied in various tasks to replace RNNs \cite{vaswani2017attention,yu2018qanet}. Self-attention mechanism is powerful to model long-term dependencies, easy for deeper architecture and parallel computing \cite{devlin2018bert}. Therefore, self-attention was introduced to model for speech recognition \cite{dong2018speech,povey2018time,salazar2019self} and achieved competitive results. 

In this paper, we propose a self-attention transducer (SA-T), an RNN-free end-to-end model for speech recognition. SA-T utilizes self-attention blocks \cite{vaswani2017attention} instead of RNNs to model long-term dependencies inside sequences. Inspired by the joint CTC-CE training \cite{zhang2018acoustic,sak2015learning}, which utilizes cross entropy (CE) loss to help CTC learn alignments and increase the stability of training, we propose an analogous method named path-aware regularization (PAR) to further improve the performance of SA-T. Path-aware regularization first constructs a possible alignment path based on forced alignment and target sequence. During training, it maximizes the probabilities of the alignment path and optimizes transducer loss function simultaneously. Self-attention mechanism requires complete sequences as inputs to model long-term dependencies, which make it difficult for SA-T to be directly applied to streaming speech recognition. Therefore, we propose a chunk-flow mechanism to alleviate this problem. Chunk-flow mechanism limits the scope of self-attention by applying a sliding-window, and stacks multiple self-attention blocks to model long-term dependencies.

All experiments are conducted on a public Mandarin Chinese dataset AISHELL-1. Our results show that the SA-T with path-aware regularization yields a 11.1\% and 21.3\% relative CER reduction compared to the plain SA-T and baseline RNN-T respectively. It also outperforms the attention-based LAS model reported in \cite{changhaoshan2019}. In addition, chunk-flow mechanism can help SA-T perform online decoding with only a little degradation of the performance.

The remainder of this paper is organized as follows. Section 2 describe the related works. Section 3 and Section 4 introduce RNN-T and our proposed self-attention transducers respectively. Section 5 presents our experimental results. The conclusions are given in Section 6. 

\section{Related Work}
Recently, there have been several works that have applied self-attention mechanism in speech recognition and achieved comparable results with traditional hybrid models \cite{dong2018speech,salazar2019self,dong2019self-attention}. Different from these, we introduce self-attention mechanism into transducer-based model. We also propose a chunk-flow mechanism to realize online decoding. Different from chunk-hopping mechanism in \cite{dong2019self-attention}, which segments an entire utterance into several overlapped chunks as the inputs, we utilize a sliding window at each layer to limit the scope of the self-attention. Chunk-flow mechanism is more analogous to the time-restricted self-attention layer \cite{povey2018time}. Both of them use sliding windows to model local dependencies between the inputs. Time-restricted self-attention layer consists of an affine component, a attention component with non-linearity function and batch normalization. Without any modification of self-attention network structure, chunk-flow mechanism just utilizes a sliding-window to limit the scope of attention and stacks multiple self-attention blocks to model the long-term dependencies.

\begin{figure*}[t]
  \centering
  \subfigure[Self-Attention Transducer]{
  \centering
  \label{fig:transducer}
  \includegraphics[width=0.25\linewidth]{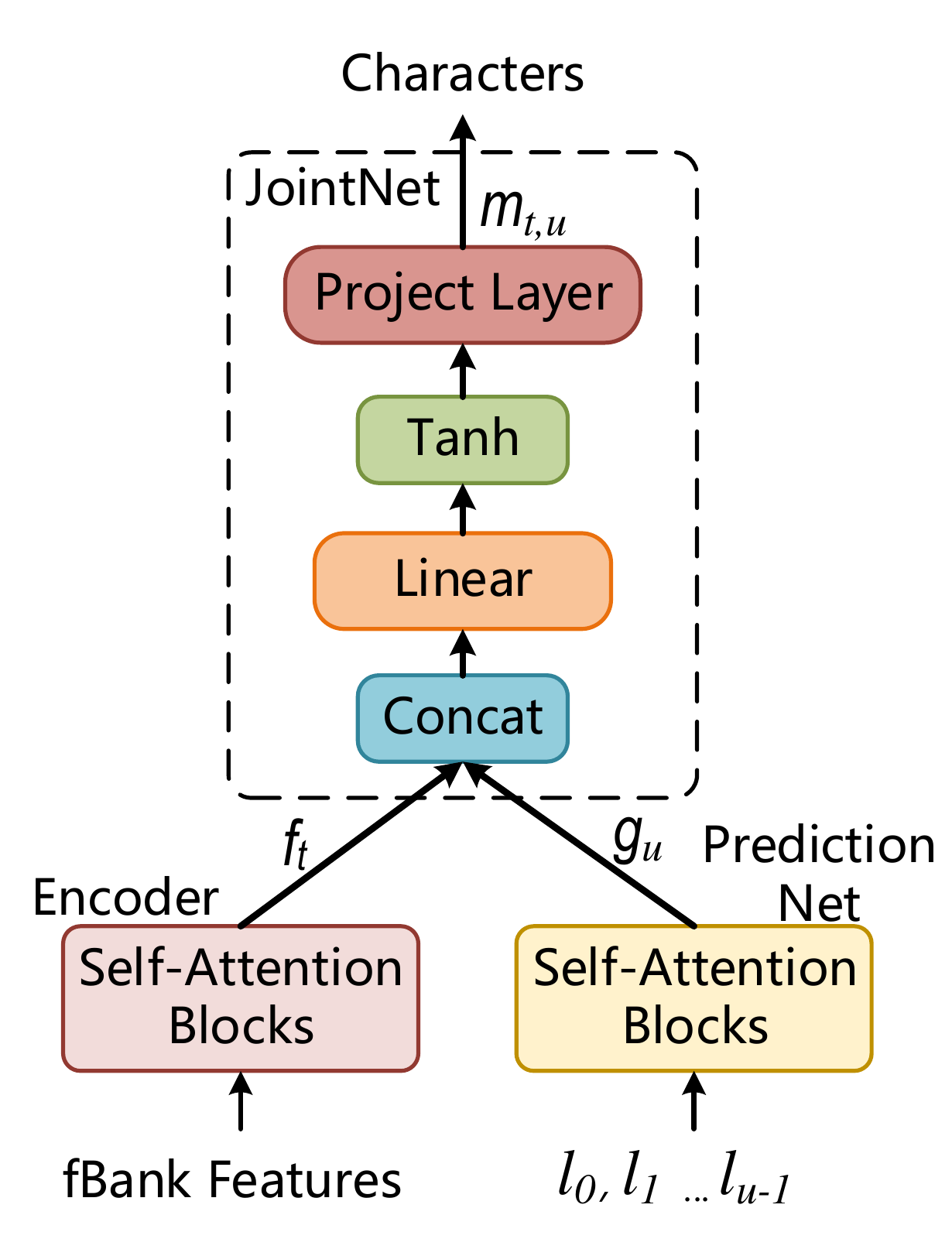}
  }
  \subfigure[Self-Attention Block]{
  \centering
  \label{fig:self-attention}
  \includegraphics[width=0.25\linewidth]{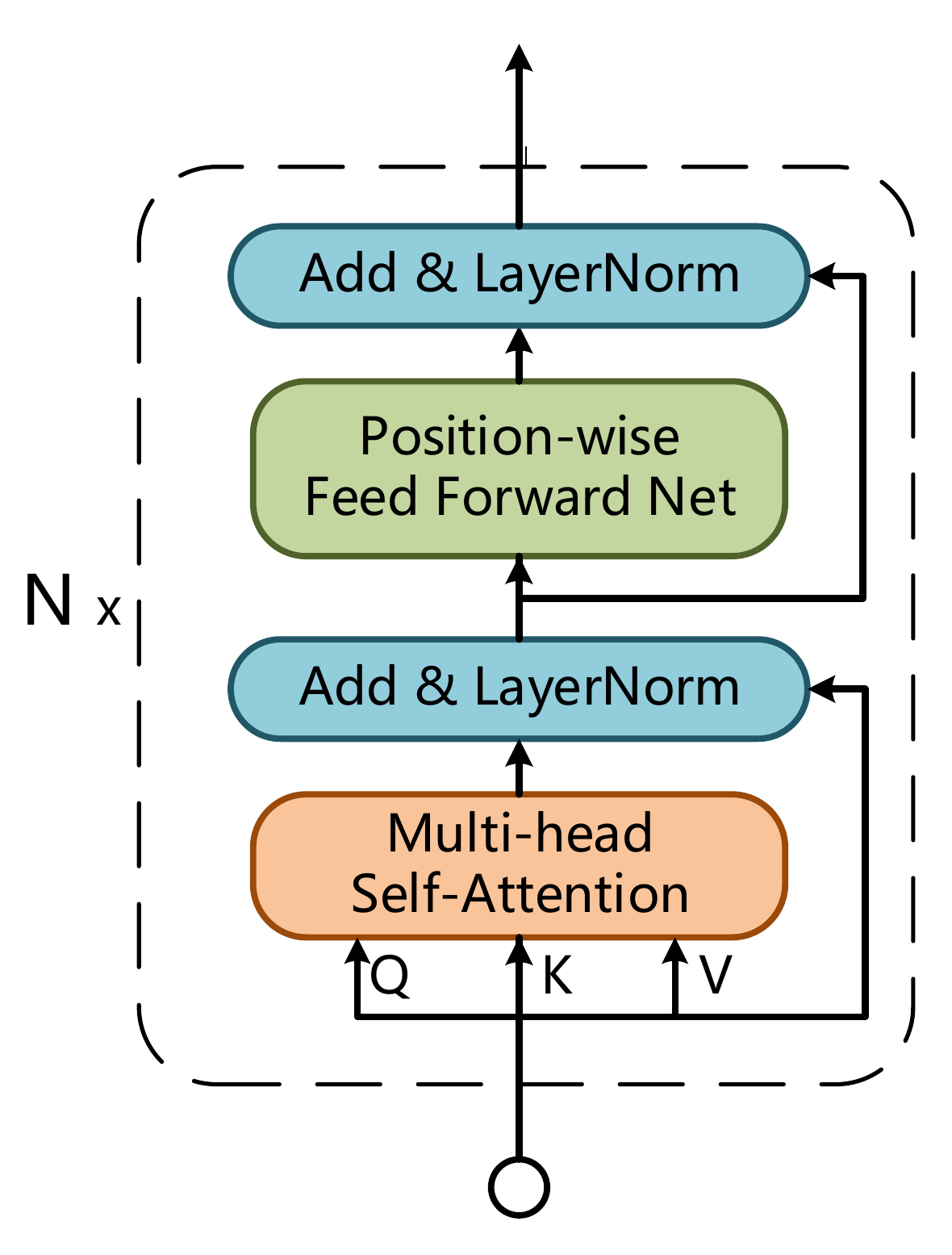}
  }
  \subfigure[Output Probability Graph]{
  \centering
  \label{fig:decoding}
  \includegraphics[width=0.3\linewidth]{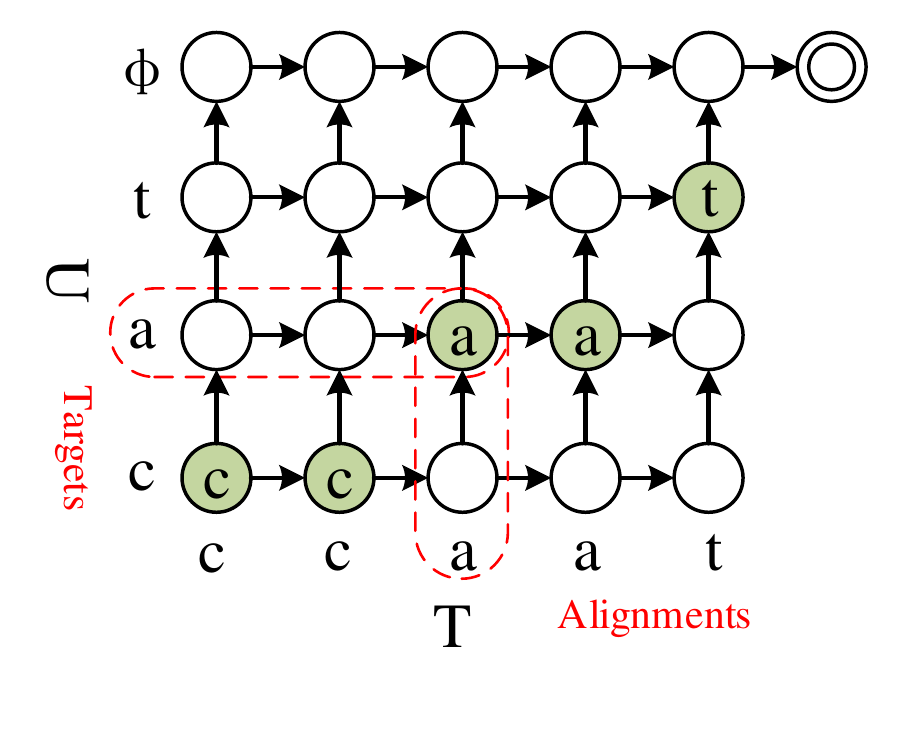}
  }
  \vspace{-5pt}
  \caption{(a) A self-attention transducer which replaces RNNs with self-attention blocks, where $f_t$, $g_u$ and $m_{t,u}$ denote the acoustic state, linguistic state and joint state respectively. (b) A basic self-attention block. (c) An output probability graph defined by $P(k|t, u)$. The horizontal axis represents time steps and the vertical axis represents label length. All possible alignment paths begin with left-down and end with the right-up double-circle ending token. The green circles denote frame-level alignment labels depicted in section 4.2.}
  \label{fig:rnntandsat}
  \vspace{-10pt}
\end{figure*}

\section{RNN-Transducer}
RNN-T consists of an acoustic encoder network, a separate language model named prediction network and a joint network \cite{graves2012sequence}. The acoustic model, which maps a input feature frame $x_t$ into a acoustic state vector $f_t$. The prediction network receives the prediction \emph{non-blank} symbol $l_{u-1}$ of the previous time step as input and computes a linguistic state vector $g_{u}$. The joint network generates a joint state $m_{t,u}$ based on the states from acoustic encoder and prediction network, and then computes the probability distribution $p(k|t,u)$ over the next output symbol ($k$ means the $k$-th token in vocabulary). 
\begin{equation}
    \label{eq:joint}
    m_{t,u} =\text{JointNet}(f_t, g_u)
\end{equation}
\begin{equation}
    \label{eq:softmax}
    p(k|t, u)=\text{Softmax}(m_{t,u})
\end{equation}
An output probability graph is illustrated in Fig. \ref{fig:decoding}. Each circle represents a posterior probability distribution estimated by joint states $m_{t,u}$. Arrows describe the direction of probability transition. Given a input feature sequence $\textbf{x}$, in order to calculate the probability $P(\textbf{y}^*|\textbf{x})$ of the corresponding target $\textbf{y}^*$, we need to sum all possible path in the output probability graph. RNN-T sums the probabilities of all possible paths by forward-backward algorithm and minimizes the negative log-likelihood of targets sequence $\mathcal{L}_{transducer}=-\text{ln}P(\textbf{y}^*|\textbf{x})$. 

\section{Self-Attention Transducer}
\subsection{Model Architecture}
To improve the computational efficiency and performance, we propose an RNN-free end-to-end model named self-attention transducer (SA-T). The SA-T replaces RNNs in encoder and prediction network with self-attention blocks \cite{vaswani2017attention} illustrated in Fig. \ref{fig:self-attention}. A self-attention block usually consists of a multi-head self-attention layer, a position-wise feed forward layer, residual connections and layer normalization.

Multi-head attention (MHA) allows the model to jointly attend to information from different positions. Each head $h_i$ is a complete self-attention component. $Q$, $K$ and $V$ represent queries, keys and values respectively. $d_k$ is the dimension of keys. $W^Q\in\mathbb{R}^{d_m\times{d_q}}$, $W^K \in \mathbb{R}^{d_m\times{d_k}}$, $W^V\in\mathbb{R}^{d_m\times{d_v}}$ and $W^O\in\mathbb{R}^{d_m\times{d_m}}$ are projection parameter matrices. 
\begin{equation}
\label{eq:self-attention}
\text{SelfAttn}(Q,K,V)=\text{softmax}(\frac{QK^T}{\sqrt{d_k}})V
\end{equation}
\begin{equation}
\begin{split}
     \text{MultiHead}(Q,K,V)&=\text{Concat}(h_1,h_2,...h_{n_h})W^O \\
     \text{where } h_i =& \text{SelfAttn}(QW_i^Q,KW_i^K,VW_i^V)
\end{split}
\end{equation}

Position-wise feed forward network (FFN) contains two linear transformations and a ReLU activation function. 
\begin{equation}
    FFN(x)=\text{max}(0,xW_1+b_1)W_2+b_2
\end{equation}
where parameters $W_1 \in \mathbb{R}^{d_m\times{d_{ff}}}$, $W_2\in \mathbb{R}^{d_{ff}\times{d_{m}}}$,, $b_1\in\mathbb{R}^{d_{m}}$ and $b_2\in\mathbb{R}^{d_{m}}$ are learned. 

Since there are no any recurrent or convolutional structures in self-attention block, we must add position encoding to the model. The sine and consine positional embedding proposed by \cite{vaswani2017attention} are applied for all the experiments in this paper. Besides, the transformer also apply residual connection and layer normalization. That is, the output of each sub-layer is $\text{LayerNorm}(x+\text{Sublayer}(x))$, where $\text{Sublayer}(x)$ may be MHA or FFN.

\subsection{Path-Aware Regularization}
Joint CTC-CE training plays a great role in assisting CTC model to coverage and learn alignments \cite{zhang2018acoustic,sak2015learning}. Considering the similarity between Transducer and CTC, we propose a path-aware regularization (PAR) to assist SA-T training.
\begin{equation}
    \label{eq:loss}
    \mathcal{L}_{joint}(\textbf{x})=\mathcal{L}_{transducer}(\textbf{x})+\beta\cdot \mathcal{L}_{par}(\textbf{x})
\end{equation}
\begin{equation}
    \label{eq:par}
    \begin{split}
        \mathcal{L}_{par}(\textbf{x}) &= -\sum_{t=0}^{T-1}\sum_{u=0}^{U-1}\sum_{k=0}^{K-1}w_{t,u}c_{t,u,k}\text{log}p(k|t,u) \\
        &\text{where } w_{t,u}=1-p(\phi|t,u)
    \end{split}
\end{equation}
where $\beta$ is a pre-set hyperparameter. $T$, $U$ and $K$ denote the length of acoustic feature sequence, the length of label sequence and the vocabulary size respectively. $c_{t,u,k}$ denotes frame-level target label. $p(\phi|t,u)$ in Eq. \ref{eq:par} denotes the probability of 'blank' label. $w_{t,u}$ is a weight used to balance regularization and transducer loss and avoid divergences in training process.

To obtain a frame-level target label $c_{t,u,k}$,
We first use Kaldi Toolkit \cite{povey2011kaldi} to generate character-level alignment sequences $A=\{a_t|0\le{t}\le{T-1}\}$ for every utterance in train set, and then record the alignment position $\textbf{c}_{t,u}$ where $a_t$ is equal to target label $l_u$ ($l_u$ belongs to target label sequences $L=\{l_u|0\le{u}\le{U-1}\}$), as depicted in Fig. \ref{fig:decoding}. $\textbf{c}_{t,u}$ is a one-hot vector, where the $k$-th element is $1$. For convenience, we replace all the silence token in the alignments with 'blank'. During training, we only compute the cross entropy loss of the alignment positions indicated by green circles in Fig. \ref{fig:decoding} and ignore other states beyond the alignment path. In other words, we first provide a reasonable alignment path for the model based on prior knowledge, and then make the model pay more attention to optimizing the probability of this path. 

\subsection{Chunk-Flow Mechanism}
Transducer-based model can decode an utterance from left to right along the acoustic feature sequences. However, because self-attention mechanism is applied in SA-T, we have to utilize a whole feature sequences as input of the model to compute attention weights. This property endues self-attention with great power to model long-term dependencies, but also prevents self-attention from being applied to model streaming sequences. Inspired by time delay neural network \cite{snyder2015time, peddinti2015time}, which can efficiently model long temporal contexts by stacking multiple time delay layers to acquire a larger receptive field, we propose a chunk-flow mechanism to improve self-attention for streaming speech recognition.

\begin{figure}[t]
  \centering
  \includegraphics[width=0.9\linewidth]{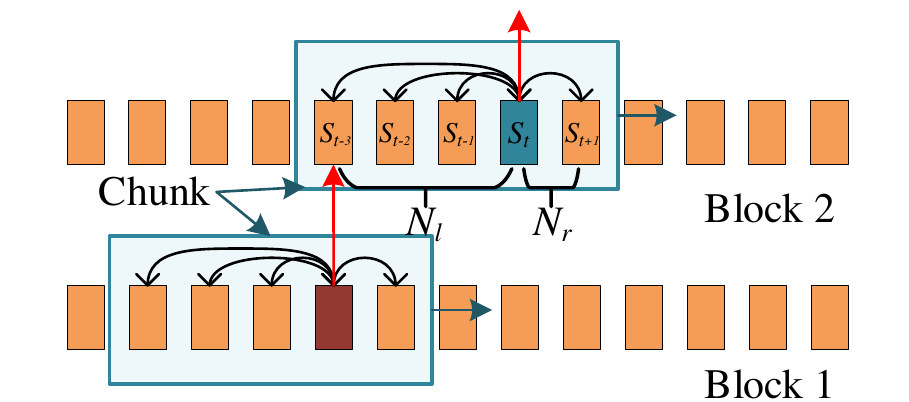}
  \caption{Chunk-Flow Mechanism. A blue box represents a chunk sliding along the time axis. Red arrows represent information flow between two layers and black arrows represent the calculation of self-attention weights.}
  \label{fig:chunk}
  \vspace{-10pt}
\end{figure}

Chunk-flow mechanism limits the scope of the self-attention blocks by utilizing a fixed-length chunk instead of a whole feature sequence as input. As illustrated in Fig. \ref{fig:chunk}, a fixed-length chunk flows along the time axis of the feature sequence. And stacking multiple self-attention blocks makes it can model longer temporal contexts without too much degradation of performance. The improved chunk-flow mechanism can be defined by the following formula.
\begin{equation}
    h_{i,t}=\sum_{\tau=t-N_l}^{t+N_r}\alpha_{i,\tau}s_{\tau}
\end{equation}
where $h_{i,t}$ denotes the $i$-th head in multi-head attention layer at time $t$, $s_\tau$ denotes the $\tau$-th vector in input sequence illustrated in Fig. \ref{fig:chunk}, $\alpha_{i,\tau}=\text{SelfAttn}(s_\tau,K,V)$ and $K=V=\text{Chunk}_\tau$. $N_l$ and $N_r$ denote the number of frames on the left and right of current time $t$ respectively. The length of a chunk in each block is equal to $N_l+N_r+1$.

\section{Experiments}
\subsection{Dataset}
In this work, all experiments are conducted on a public Mandarin speech corpus AISHELL-1\footnote{http://www.openslr.org/13/} \cite{DBLP:journals/corr/abs-1709-05522}. The training set contains about 150 hours of speech (120,098 utterances) recorded by 340 speakers. The development set contains about 20 hours (14,326 utterances) recorded by 40 speakers. And about 10 hours (7,176 utterances) speech is used to be test set. The speaker of different sets are not overlap.

\subsection{Experimental Setup}
For all experiments, we use 40-dimension Mel-filter bank coefficients (Fbank) features computed on 25ms window with 10ms shift. Each feature is re-scaled to have zero mean and unit variance for each speaker. Similar to \cite{pundak2016lower}, at the current frame, $t$, these features are stacked with 3 frames to the left and 1 frames to the right, and then downsampled to a 30ms frame rate. We chose 4232 characters (including 'blank' and 'unk' labels) as model units. At training time, all utterances are sorted by length. We use PyTorch \cite{paszke2017automatic} for modeling and Kaldi \cite{povey2011kaldi} for data preparation.

For the baseline RNN-T model, we utilize a 4-layer bi-directional LSTM with 320 hidden units each direction as encoder, a 2-layer uni-directional LSTM with 512 hidden units as prediction network. The joint network combines acoustic and linguistic states, and projects the output of non-linear activation function to softmax layer. Both input and project linear layers in the joint network have 512 hidden units. The RNN-T model is optimized by stochastic gradient decent (SGD) optimizer with learning rate 0.001 and momentum 0.9. If it has no improvement, then halve the learning rate and repeat this process until the learning rate is less than $1\times10^{-6}$. 

All the SA-T models use 6 self-attention blocks as encoder and 4 self-attention blocks as prediction network. We take ($d_{m}$, $n_{h}$, $d_{ff}$)=($512$, $8$, $1024$). Our proposed SA-T model is optimized in the same method as \cite{salazar2019self}. And we set warm up steps to 8000 and the factor $\lambda$ of learning rate to 0.5.

During decoding, we use beam search with width of 5 \cite{graves2012sequence} for all the experiments. A character-level 5-gram language model from training text, is integrated into beam searching by shallow fusion \cite{kannan2018an}. The weight of language model is set to 0.2 for all the experiments.
\begin{table}[th]
  \caption{Comparisons of the SA-T with different weights.}
  \label{tab:norm}
  \centering
  \begin{tabular}{c|c|cc|cc}
    \toprule
    \textbf{Model} & $\bm{\beta}$ & \textbf{Dev} & \textbf{Gain} & \textbf{Test} & \textbf{Gain} \\
    \hline
    \makecell[l]{SA-T} & 0.0 & 9.21 & - & 10.46 & - \\
    \hline
    \multirow{6}*{\makecell[r]{ + Path-Aware}} & 1.0 & 8.56 & 7.1\% & 9.65 & 7.8 \% \\
     & 2.0 & 8.51 & 7.6\% & 9.63 & 8.0 \%\\
     & 5.0 & 8.33 & 9.6\% & 9.48 & 9.4 \% \\
     & 10.0 & \textbf{8.30} & \textbf{9.9\%} & \textbf{9.30} & \textbf{11.1\%} \\
     & 15.0 & 8.31 & 9.8\% & 9.41 & 10.4\% \\
     & 20.0 & 8.35 & 9.3\% & 9.43 & 9.8\% \\
    \bottomrule
  \end{tabular}
  \vspace{-10pt}
\end{table}
\vspace{0pt}

\begin{figure}[t]
  \centering
  \includegraphics[width=\linewidth]{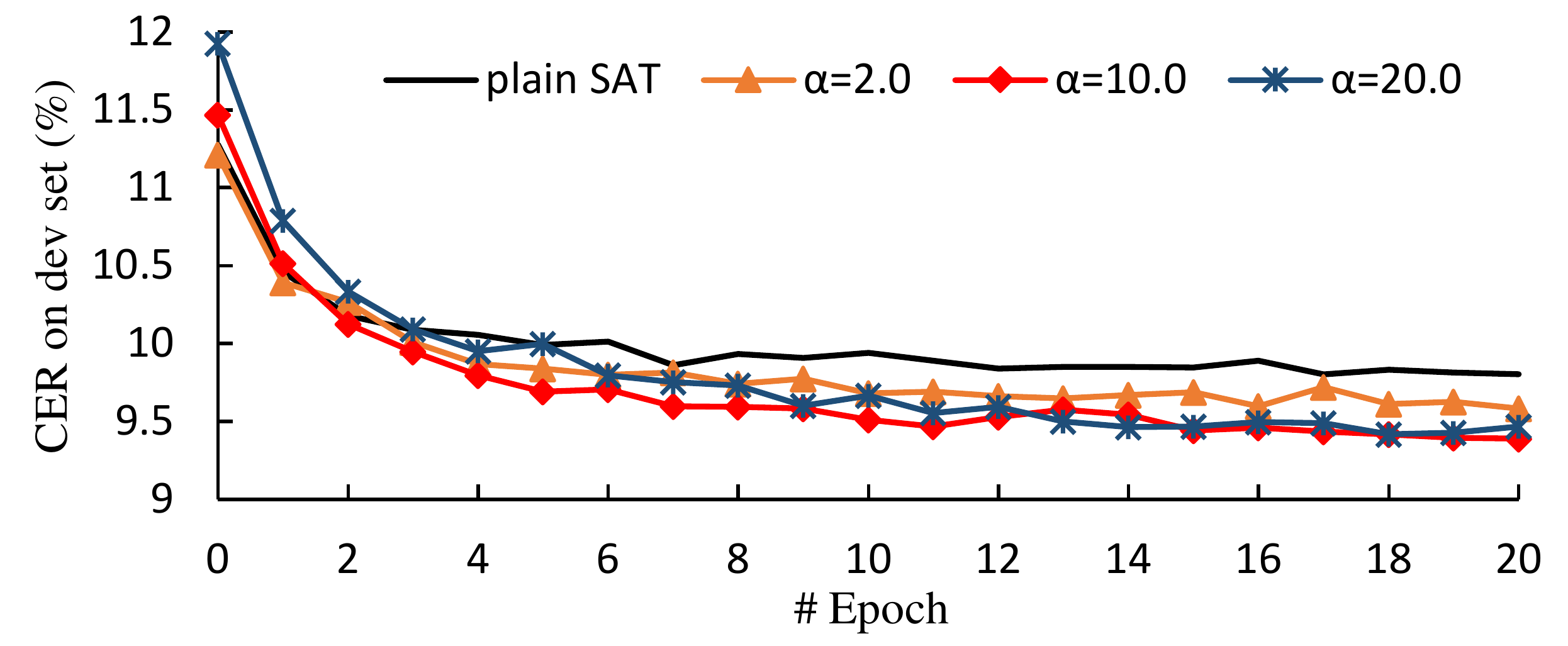}
  \vspace{-10pt}
  \caption{Comparison of various learning curves of the SA-T with path-aware regularization}
  \label{fig:comparasion}
  \vspace{-10pt}
\end{figure}
\begin{table}[t]
  \caption{Comparisons of the SA-T with different length contexts.}
  \vspace{-5pt}
  \label{tab:window}
  \centering
  \begin{tabular}{c|cccc}
    \toprule
    \textbf{Model} & $\bm{N_l}$ & $\bm{N_r}$ & \textbf{Dev} & \textbf{Test} \\
    \hline
    \multirow{6}*{\makecell[l]{SA-T \\ \\ \\ + Chunk-Flow Mechanism \\ \\ \\}} & - & - & 9.21 & 10.46 \\ 
    \hline
    & 5 & 2 & 11.17 & 12.51 \\
    & 10 & 2 & 11.15 & 12.42 \\
    & 10 & 5 & 10.43 & 11.80 \\
    & 20 & 5 & 10.26 & 11.57 \\
    & 20 & 10 & 9.81 & 11.06 \\
    \hline
    \makecell[l]{+ Path-Aware ($\beta=10$)}& 20 & 10 & \textbf{8.58} & \textbf{9.80} \\
    \bottomrule
  \end{tabular}
  \vspace{-15pt}
\end{table}
\vspace{0pt}
\subsection{Path-Aware Regularization}
We train SA-Ts using path-aware regularization with various weights $\beta$. if $\beta$ is equal to 0, then the model is equivalent to plain SA-T. Table \ref{tab:norm} demonstrates the relationship between different weights and the performance of a model. We find that SA-T can achieve a CER of 10.46\% on test set and the SA-Ts with regularization have consistently outperformed the plain SA-T. When $\beta$ is 10, the model can achieve up to 11.1 \% relative CER reduction compared to the plain SA-T. Learning curves in Fig. \ref{fig:comparasion} indicate that regularization can assist SA-T to accelerate convergence and improve the performance. The results also indicate that SA-T can achieve better performance when the weight of regularization is great than 5. We suppose that a big wight can force the model to learn alignments better.

\begin{figure}[t]
  \centering
  \subfigure[SA-T]{
  \centering
  \label{fig:alignment1}
  \includegraphics[width=0.8\linewidth]{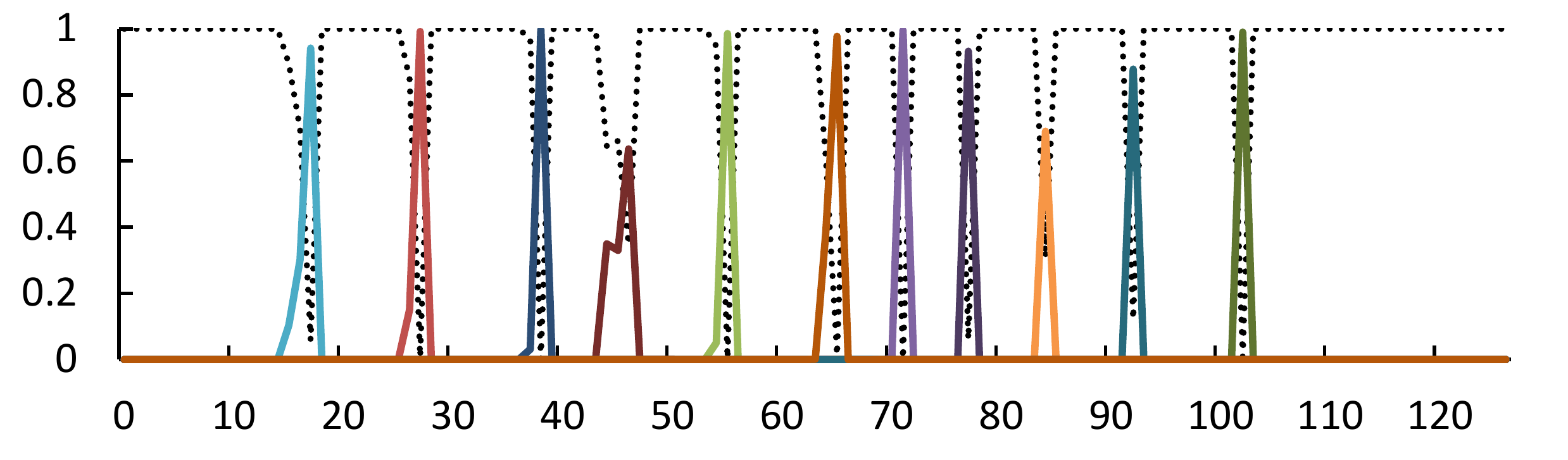}
  }
  \subfigure[Alignment Boundary]{
  \centering
  \label{fig:alignment}
  \includegraphics[width=0.8\linewidth]{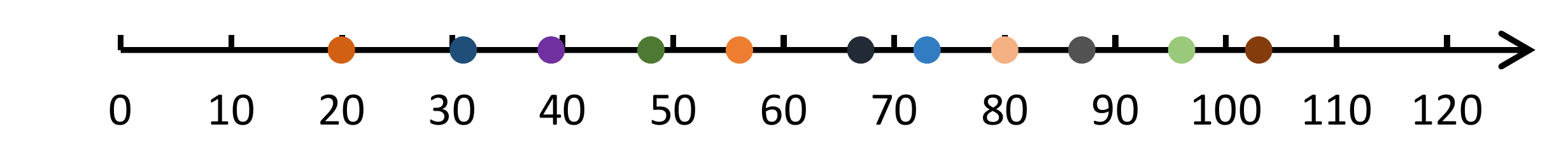}
  }
  \subfigure[SA-T with Path-Aware Regularization ($\beta=10.0$)]{
  \centering
  \label{fig:alignment2}
  \includegraphics[width=0.8\linewidth]{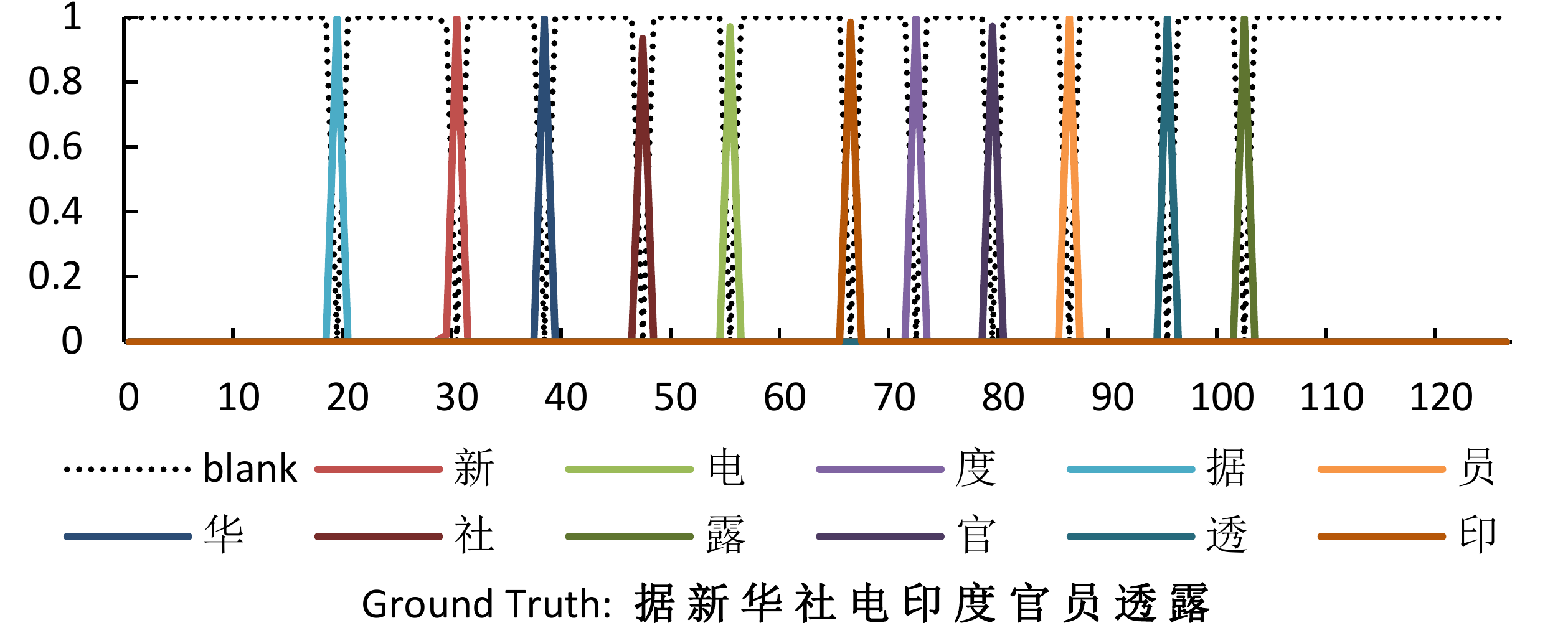}
  }
  \vspace{-10pt}
  \caption{Label posteriors estimated by SA-Ts on test set during greedy search decoding. (a)Plain SA-T. (b) Alignment boundary generated by Kaldi toolkit. (c) SA-T training with path-aware regularization. }
  \label{fig:align}
\end{figure}

To further research the effects of our proposed path-aware regularization, we capture the probability distribution estimated by the SA-T models during greedy search decoding on test set, as illustrated in Fig. \ref{fig:alignment2}. It is clear that path-aware regularization can help the model learn a better alignment and sharper probability distribution. As a result, a SA-T with path-aware regularization tends to achieve a better performance than the plain SA-T.
\vspace{-5pt}
\subsection{Chunk-Flow Mechanism}
We compare the effects of different length chunks on model performance, as illustrated in Table \ref{tab:window}. It is easy to observe that the longer contexts be modeled, and the better performances model have, which is consistent with our common sense. Chunk-flow SA-T with 20 left-context frames and 10 right-context frames has the best performance that is competitive to baseline RNN-T. The results indicate that chunk-flow mechanism can model long-term dependencies by stacking multiple blocks and implement online decoding without too much degradation of performance. Furthermore, we also train a chunk-flow SA-T with path-aware regularization ($\beta=10$), and observe that the model get 12.5\% and 11.4\% relative CER reduction on dev and test set respectively compared to the one without regularization. It proves the effectiveness of path-aware regularization again.

\vspace{0pt}
\subsection{Compare the SA-T and other models}
Table \ref{tab:base} compares SA-T with other models.  Compared to the baseline RNN-T, the SA-T achieves 8.2\% and 9.5\% relative CER reduction on dev and test set respectively. And it also outperforms the attention-based LAS model reported in \cite{changhaoshan2019}. We suppose that more trainable parameters, deeper network structure and more effective method to model long-term dependencies make the SA-T obtain a good performance. Furthermore, we notice that the SA-T with path-aware regularization can achieve a 21.3\% relative CER reduction over the baseline RNN-T.

\begin{table}[t]
  \caption{Comparison of SA-T and other models.}
  \vspace{-5pt}
  \label{tab:base}
  \centering
  \begin{tabular}{c|cc}
    \toprule
    \textbf{Model} & \textbf{Dev} & \textbf{Test} \\
    \hline
    \makecell[l]{LAS \cite{changhaoshan2019}} & - & 10.56 \\
    \makecell[l]{RNN-T(baseline)} & 10.13  & 11.82 \\
    \makecell[l]{SA-T (proposed)} & 9.21  & 10.46 \\
    + Path-Aware (proposed) & \textbf{8.30}  & \textbf{9.30} \\
    \bottomrule
  \end{tabular}
  \vspace{-10pt}
\end{table}
\vspace{-10pt}
\section{Conclusions}
In this work, we propose a self-attention transducer, which replaces RNNs with self-attention blocks. Self-attention transducer outperforms the baseline RNN-T on AISELL-1. A path-aware regularization, similar to cross entropy regularization, is proposed to improve the stability of SA-T training and performance. The results show that SA-T with regularization has a great improvement on the plain SA-T. We also utilize a general chunk-flow mechanism to achieve online decoding. During decoding, we notice that the model easily predicts a character with similar pronunciation. In the future, we will explore how to improve the language modeling ability.  

\section{Acknowledge}
This work is supported by the National Key Research \& Development Plan of China (No.2018YFB1005003), the National Natural Science Foundation of China (NSFC) (No.61425017, No.61831022, No.61771472, No.61603390), the Strategic Priority Research Program of Chinese Academy of Sciences (No.XDC02050100), and Inria-CAS Joint Research Project (No.173211KYSB20170061).

\bibliographystyle{IEEEtran}

\bibliography{mybib}

\end{document}